\documentclass[english,aps,pre,a4paper,preprint,groupedaddress,floatfix,longbibliography]{revtex4}
\usepackage{bm}
\usepackage{amssymb,amsmath}
\usepackage{graphicx}
\usepackage{esint}
\usepackage{hyperref}
\begin{document}

\title{Beyond the Knudsen number: assessing thermodynamic non-equilibrium in gas flows}

\author{Jianping Meng} \thanks{These authors contribute equally} \email{jianping.meng@strath.ac.uk}
\affiliation{Department of Mechanical and Aerospace Engineering, University of
Strathclyde, Glasgow G1 1XJ, UK}

\author{Nishanth Dongari} \thanks{These authors contribute equally}\email{nishanth.dongari@strath.ac.uk}
\affiliation{Department of Mechanical and Aerospace Engineering, University of
Strathclyde, Glasgow G1 1XJ, UK}

\author{Jason M Reese}  \email{jason.reese@strath.ac.uk}
\affiliation{Department of Mechanical and Aerospace Engineering, University of
Strathclyde, Glasgow G1 1XJ, UK}

\author{Yonghao Zhang} \thanks{To whom the correspondence should be addressed}\email{yonghao.zhang@strath.ac.uk}

\affiliation{Department of Mechanical and Aerospace Engineering, University of
Strathclyde, Glasgow G1 1XJ, UK}

\begin{abstract}
For more than 150 years the Navier-Stokes equations for thermodynamically
quasi-equilibrium flows have been the cornerstone of modern computational fluid
dynamics that underpins new fluid technologies. However, the applicable
regime of the Navier-Stokes model in terms of the level of thermodynamic
non-equilibrium in the local flowfield is not clear especially for hypersonic and low-speed micro/nano flows. Here, we re-visit the
Navier-Stokes model in the framework of Boltzmann statistics, and propose a new
and more appropriate way of assessing non-equilibrium in the local flowfield,
and the corresponding appropriateness of the Navier-Stokes model. Our
theoretical analysis and numerical simulations confirm our proposed method.
Through molecular dynamics simulations we reveal that  the commonly-used Knudsen
number, or a parametric combination of Knudsen and Mach numbers, may not be
sufficient to accurately assess the departure of flowfields from equilibrium,
and the applicability of the Navier-Stokes model.

\end{abstract}

\maketitle

The Navier-Stokes equations are the most widely-used
model for fluid dynamics. Their impact is far-ranging, e.g. weather forecasting,
modern transport system design including aeroplanes, cars, and ships, and
energy generation from wind turbines. The fundamental assumptions for the
Navier-Stokes model are that the fluid is a continuous medium, and the flow is
close to thermodynamic equilibrium.

For a gas, the Navier-Stokes equations can be regarded
as a first order approximate solution to the Boltzmann equation, in terms of the Knudsen number $Kn$ (the ratio of the molecular mean free path $\lambda$ to the system characteristic length $L$) \cite{Chapman1953}.
The Boltzmann equation assumes both binary collisions between gas molecules
and molecular chaos, so a single particle velocity distribution function can
be employed. The Chapman-Enskog approach to the Boltzmann
equation assumes  
\begin{equation}
f=f^{(0)}+f^{(1)}+f^{(2)}+\cdots+f^{(\alpha)}+\cdots, \label{eq:fsol}
\end{equation}
where the distribution functions $f^{(\alpha)}$ can be asymptotically obtained from
the Boltzmann equation. The Maxwell-Boltzmann equilibrium distribution,
\begin{equation}
f^{eq}=\frac{\rho}{(2\pi RT)^{3/2}}\exp\left[-\frac{\varsigma^{2}}{2RT}\right],
\end{equation}
is the zeroth-order solution $f^{(0)}$, and leads to the Euler fluid equations.
Here, $\rho$ denotes the gas density, $T$ the temperature, $R$ the
gas constant, and $\bm \varsigma$ the peculiar velocity of molecules,
which is $\bm \xi-\bm u$ where $\bm \xi$ represents the molecular
velocity and $\bm u$ is the macroscopic fluid velocity. In Eq.(\ref{eq:fsol}), $f^{(1)}$
provides a non-equilibrium correction of the order of $Kn$. To recover the Navier-Stokes equations, 
\begin{equation}
f^{(1)}=f^{eq}\left[\left(\frac{\sigma_{ij}\varsigma_{<i}\varsigma_{j>}}{2pRT}\right)+\frac{2q_{i}\varsigma_{i}}{5pRT}\left(\frac{\varsigma^{2}}{2RT}-\frac{5}{2}\right)\right],  \label{eq:fNS}
\end{equation}
where, $p$ is the gas pressure, and the shear stress $\sigma_{ij}$
and the heat flux $q_{i}$ are related to the following first-order
gradients of velocity and temperature,
\begin{equation}
\sigma_{ij}=-2\mu\frac{\partial u_{<i}}{\partial x_{j>}},  \quad q_{i}=-\kappa\frac{\partial T}{\partial x_{i}},
\end{equation}
where $\mu$ and $\kappa$ denote the viscosity and thermal conductivity.
We can continue the series to obtain $\alpha$-order corrections to the distribution function in terms of the Knudsen number.

The Chapman-Enskog technique indicates that the Euler equations are appropriate for
thermodynamically equilibrium flows with $Kn=0$, while the
Navier-Stokes equations are valid for linear departures from equilibrium where
the Knudsen number is close to zero. In high altitude applications, such as
spacecraft re-entry into planetary atmospheres, or vacuum applications, low-pressure chemical processes, and micro/nano devices, the flows can be highly non-equilibrium
and the Navier-Stokes equations fail to provide an adequate description. However, the Knudsen number
alone may not be sufficient to describe the level of non-equilibrium in the flowfield,
and to assess whether the Navier-Stokes equations are applicable or not.
Both simulation data \cite{Macrossan} and theoretical analysis \cite{Tsien1946} indicate
that the level of non-equilibrium is also strongly influenced by the Mach number. 

Two types of breakdown criteria have often been used for indicating when the Navier-Stokes equations are appropriate (\cite{wang:91,2009RSPSA.465.1581L,Macrossan,Garcia1998,boyd:210}
and references therein). The first type is the local Knudsen number,
${\lambda}/{\phi}|{d\phi}/{dx}|$, where $\phi$ denotes
a macroscopic flow quantity (typically density, temperature or pressure).
The other type is the product of local Mach and Knudsen numbers, or
its equivalent \cite{Macrossan}. The first criterion indicates that the Knudsen number has
to be small for the Navier-Stokes model to be valid. The second criterion requires the product
of $Ma$ and $Kn$ to be small. However, the latter
is not appropriate for the low Mach number flows typically occurring in
micro/nano-devices, as the evidence is that the Navier-Stokes
equations fail for a small $KnMa$ but moderate $Kn$
\cite{Hadjiconstantinou2006}. 

The inconsistency in these two types of criteria
raises a fundamental question: is there a better way to assess the level of thermodynamic
non-equilibrium in the local flowfield and the appropriateness of the
Navier-Stokes equations? As the Navier-Stokes equations are
the cornerstone of modern computational fluid dynamics (CFD), their validity range as a model should be clearly defined. Here, we address this fundamental issue, aiming to accurately assess the level of non-equilibrium of the local flowfield, and so redefine the applicability regime of the Navier-Stokes equations.

Instead of using Knudsen and Mach numbers defined by macroscopic flow properties, the fundamental direct evidence for the level of thermodynamic non-equilibrium in the local flowfield resides in the distribution function itself. The distribution
function may be considered in two parts: the equilibrium and the non-equilibrium components, i.e. 
\begin{equation}
f=f^{eq}+f^{neq}.
\end{equation} 
In order to consider the validity of the Navier-Stokes equations, the non-equilibrium
part $f^{neq}$ may be further divided into two components, i.e. 
\begin{equation}
f^{neq}=f^{(1)}+f^{(H)},
\end{equation}
where $f^{(1)}$ is the first order non-equilibrium correction
(at the Navier-Stokes level) which is given by Eq.(\ref{eq:fNS}), and $f^{(H)}$ is the higher-order non-equilibrium correction beyond the Navier-Stokes level. So, the distribution function can be split into three components:
\begin{equation}f=f^{eq}+f^{(1)}+f^{(H)}.\end{equation} If the non-equilibrium part $f^{neq}$
is negligible in comparison to $f^{eq}$, then the flow is in equilibrium, and
the Euler equations can be recovered from the Boltzmann equation. Only
when $f^{(1)}\gg f^{(H)}$ are the Navier-Stokes equations recovered from the Boltzmann equation. Using the distribution function directly we can thereby assess when the Navier-Stokes
equations are valid, which is not only physically sound but also practical as
many computational methods provide information on the distribution
function during simulations, e.g. direct solution of the Boltzmann equation \cite{yensmaf},
the lattice Boltzmann method \cite{Meng2011a}, the direct simulation Monte Carlo method \cite{Bird1978}, and molecular dynamics \cite{Dongari2011}.

To evaluate how far the flowfield is away from equilibrium, we introduce
a parameter $C_{0}$ to describe the departure from local equilibrium:

\begin{equation}
\begin{split}
C_{0} &
 = 
\sqrt{\frac{\int(f^{neq})^{2}d\bm{\xi}}{\int(f^{eq})^{2}d\bm{\xi}}} =  \sqrt{
\frac
{\int(f-f^{eq})^{2}d\bm{\xi}}{\int(f^{eq})^{2}d\bm{\xi}}}\\
 & = \frac{\sqrt{8}\pi^{
3/4 }(RT)^{3/4}}{\rho}\sqrt{\int(f-f^{eq})^{2}d\bm{\xi}},\label{eq:c0}
\end{split}
\end{equation}
which is a relative error of $f$ to the Maxwellian
$f^{eq}$. Similarly, a parameter $C_{1}$ can be introduced to
describe how far the flowfield is away from the Navier-Stokes regime: 

\begin{equation}
C_{1}=\sqrt{\frac{\int(f^{(H)})^{2}d\bm{\xi}}{\int(f^{(1)})^{2}d\bm{\xi}}}=\sqrt{\frac{\int(f-f^{eq}-f^{(1)})^{2}d\bm{\xi}}{\int(f^{(1)})^{2}d\bm{\xi}}}.\label{eq:c1}
\end{equation}
Here, $C_{1}$ is a direct indicator of the relative error introduced by using a Navier-Stokes model on the flowfield. Together, $C_{0}$ and $C_{1}$ provide both an accurate assessment
of the level of non-equilibrium in the local flowfield, and an indication of the
appropriateness of using the Navier-Stokes equations. 

In the following section, we show quantitatively why the commonly used Knudsen and Mach numbers fail as local flowfield indicators, and how they are related to $C_{0}$ and $C_{1}$.
We use nonlinear shear-driven Couette flows as examples, where the two
plates are moving with a speed of $U_w$ in opposite directions with
their temperatures set to $T_0$. To obtain
accurate gas molecular velocity distribution functions, we perform molecular
dynamics (MD) simulations using the OpenFOAM code that includes the MD routines
implemented by Macpherson and Reese \emph{et al.} \cite{Macpherson2008}. Monatomic
Lennard-Jones argon molecules are simulated \cite{Dongari2011}, and initially
the molecules are spatially distributed in the domain of interest with
a random Gaussian velocity distribution corresponding to an initially
prescribed gas temperature. They are then allowed to relax through collisions until reaching a steady state before we take measurements. To achieve
a smooth velocity distribution function, molecular velocity samples
are then taken in every time step (0.001 $\tau$, where $\tau=\sqrt{md^{2}/\epsilon}$, with $m$ being molar mass, $d$, the diameter of gas molecules, and $\epsilon$ being related to the interaction strength of the molecules)
for a total run time of at least 30000 $\tau$ (in the extreme
rarefied and high speed flow case below, up to 100000 $\tau$). We use 83500 molecules in each simulation, and assume diffuse gas molecule/wall interactions. 

\begin{figure*}
\begin{center}
\includegraphics[width=.48
\textwidth]{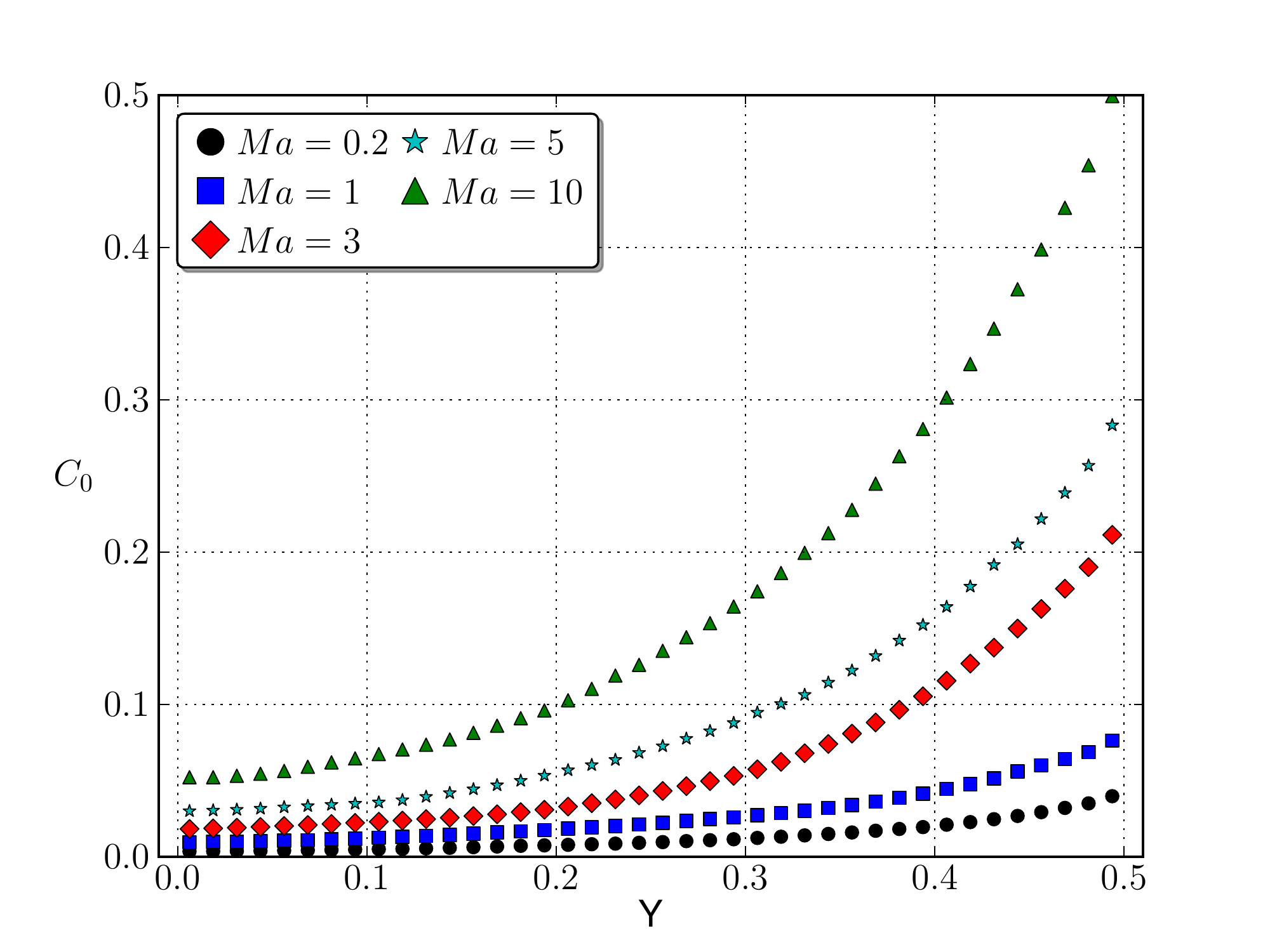}\includegraphics[width=.48 \textwidth]{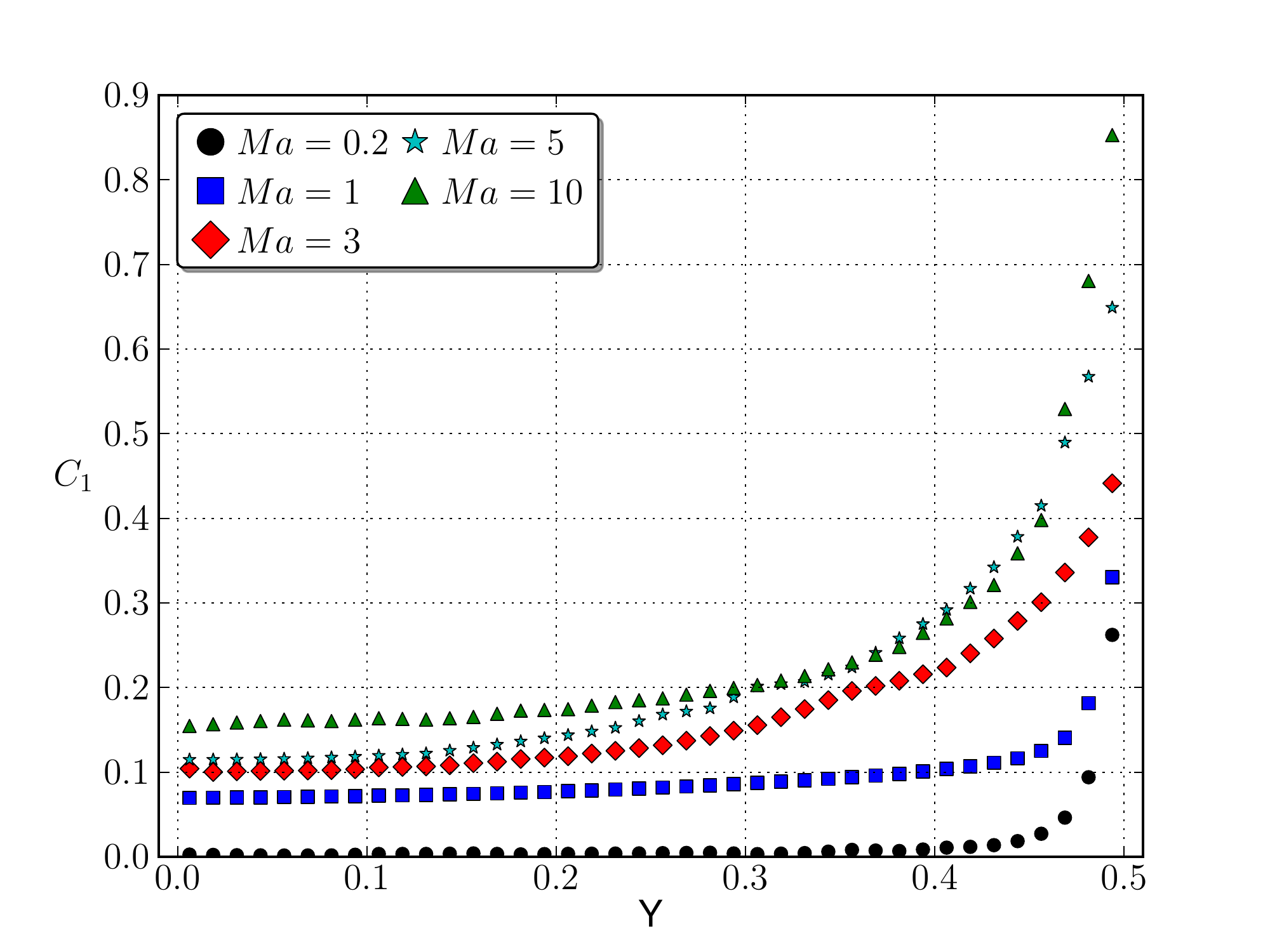}

\caption{Half-channel profiles of $C_{0}$ and $C_{1}$ at
  $Kn=0.01$\label{fig:Proc0c1kn001}, for various $Ma$. The top ($Y=0.5$) and bottom ($Y=-0.5$)
  plates are moving with speeds $U_w$ in opposite directions.}
\end{center}
\end{figure*}

\begin{figure*}
\begin{center}
\includegraphics[width=.48
\textwidth]{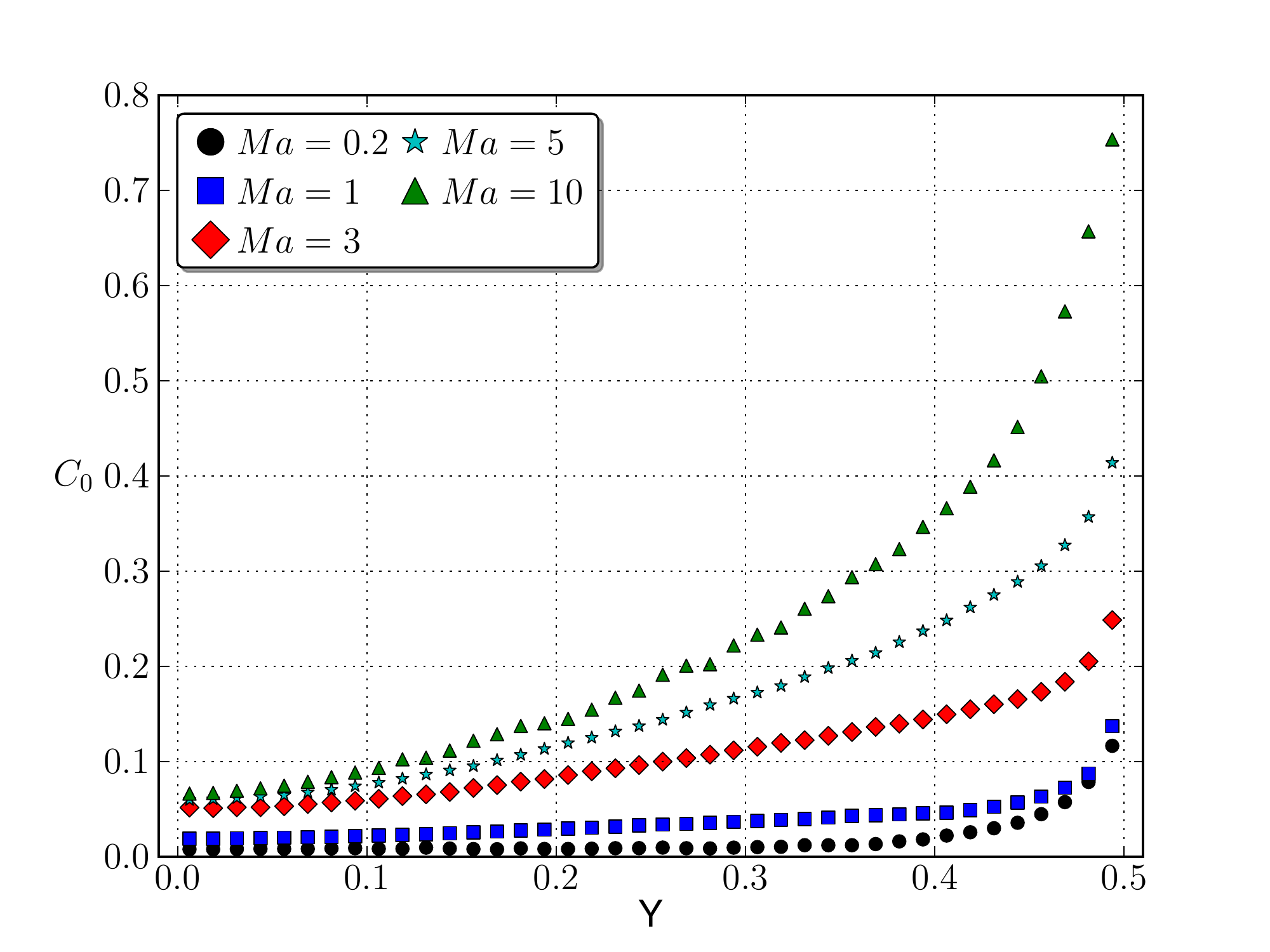}\includegraphics[width=.48 \textwidth]{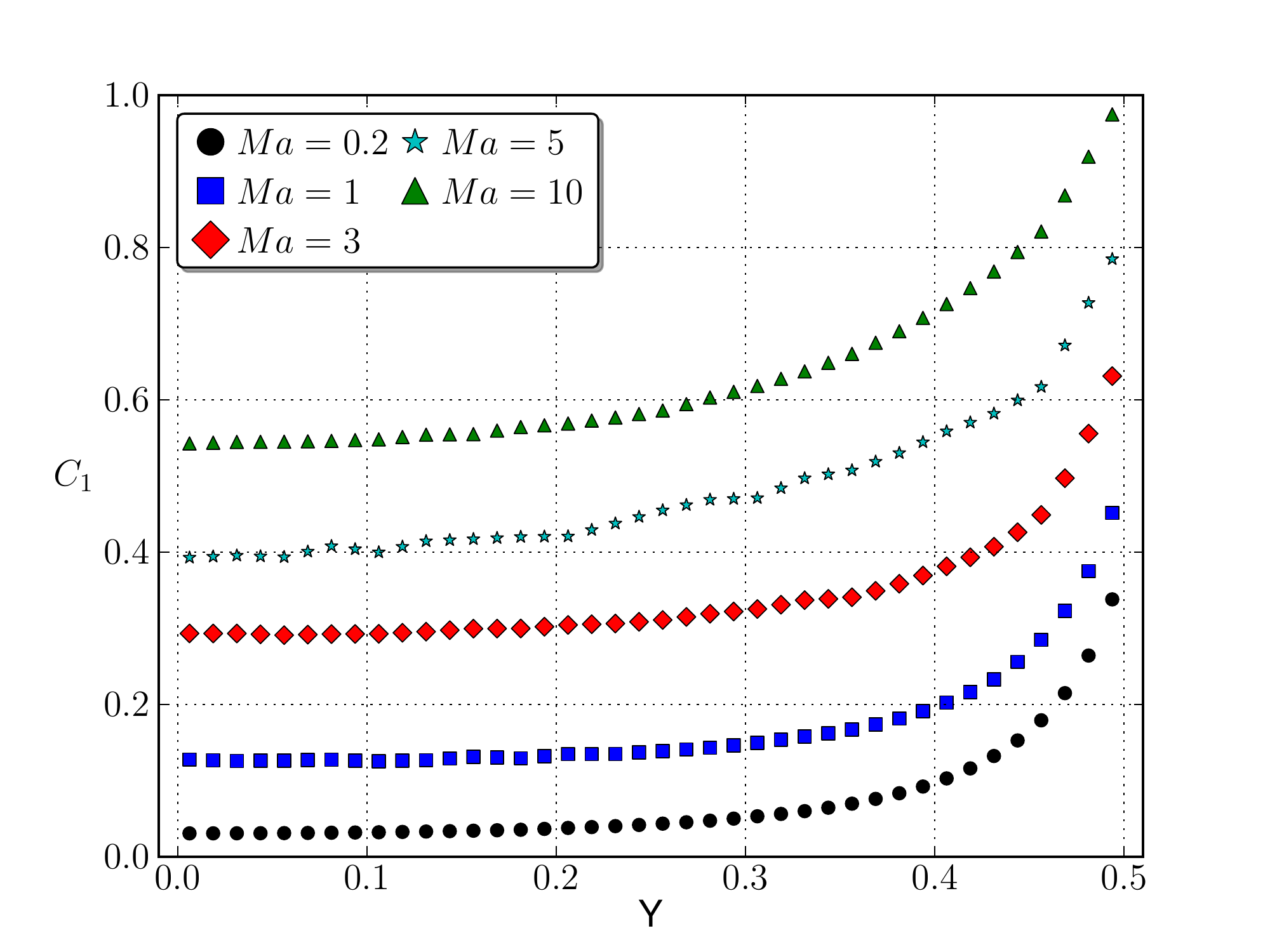}

\caption{Half-channel profiles of $C_{0}$ and $C_{1}$ at
  $Kn=0.1$,\label{fig:Proc0c1kn01} for various $Ma$.The top ($Y=0.5$) and bottom ($Y=-0.5$)
  plates are moving with speeds $U_w$ in opposite directions.}
\end{center}
\end{figure*}

\begin{figure*}
\begin{center}
\includegraphics[width=.48
\textwidth]{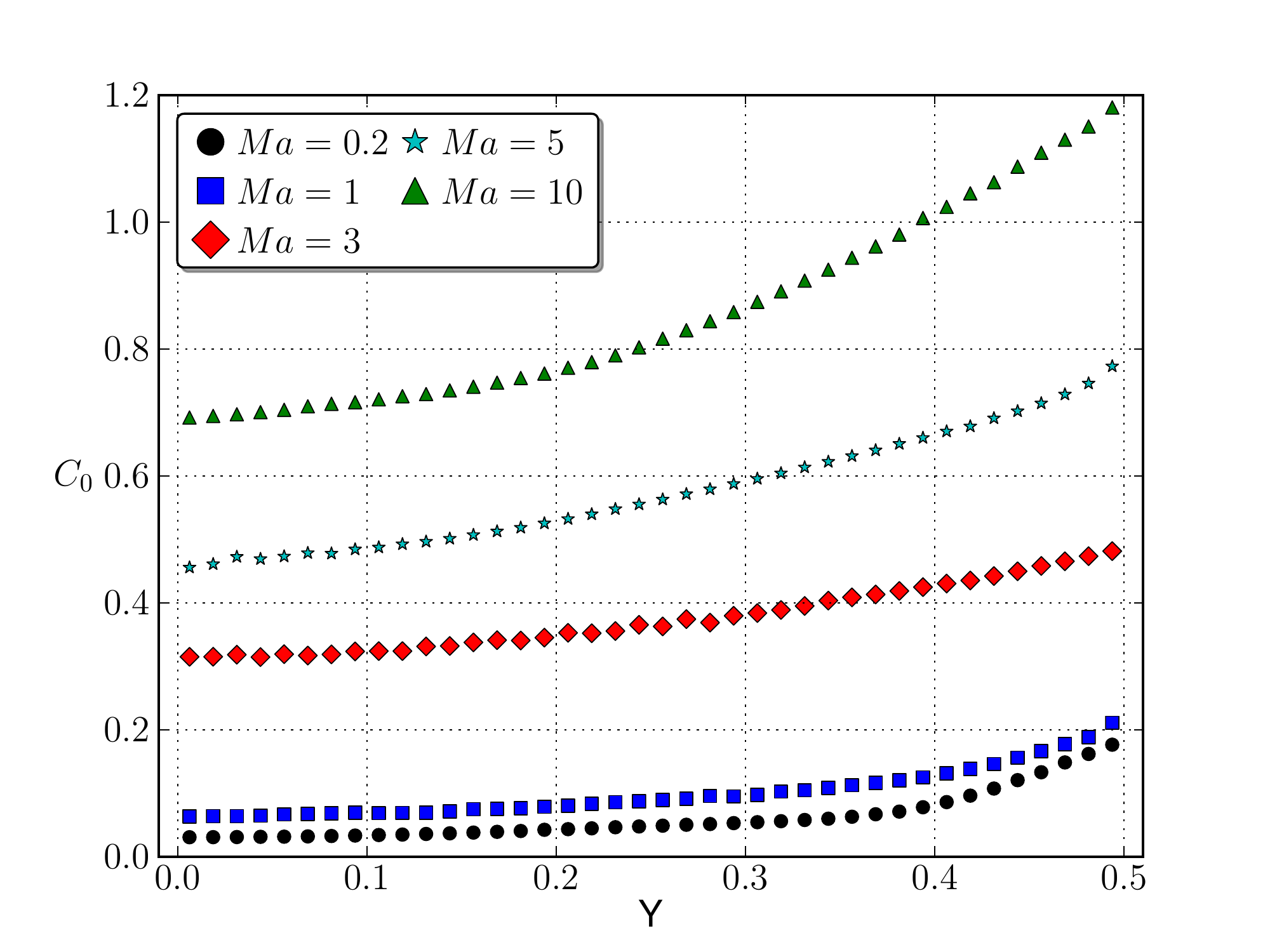}\includegraphics[width=.48
\textwidth]{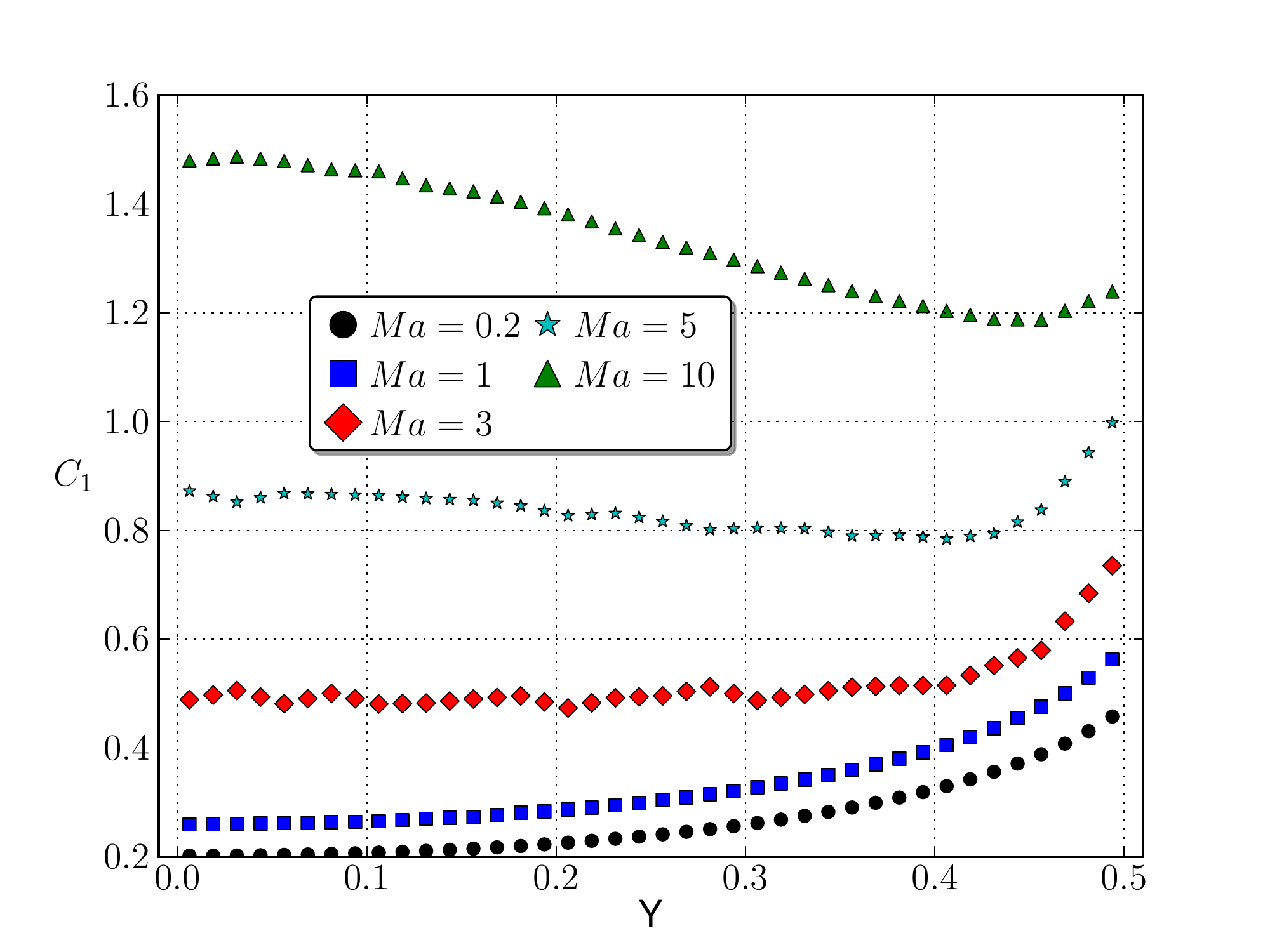}\caption{ Half-channel profiles of $C_{0}$ and $C_{1}$
at $Kn=0.5$, for various $Ma$. \label{fig:Proc0c1kn05}}
\end{center}
\end{figure*}

\begin{figure*}
\begin{center}
\includegraphics[width=.48
\textwidth]{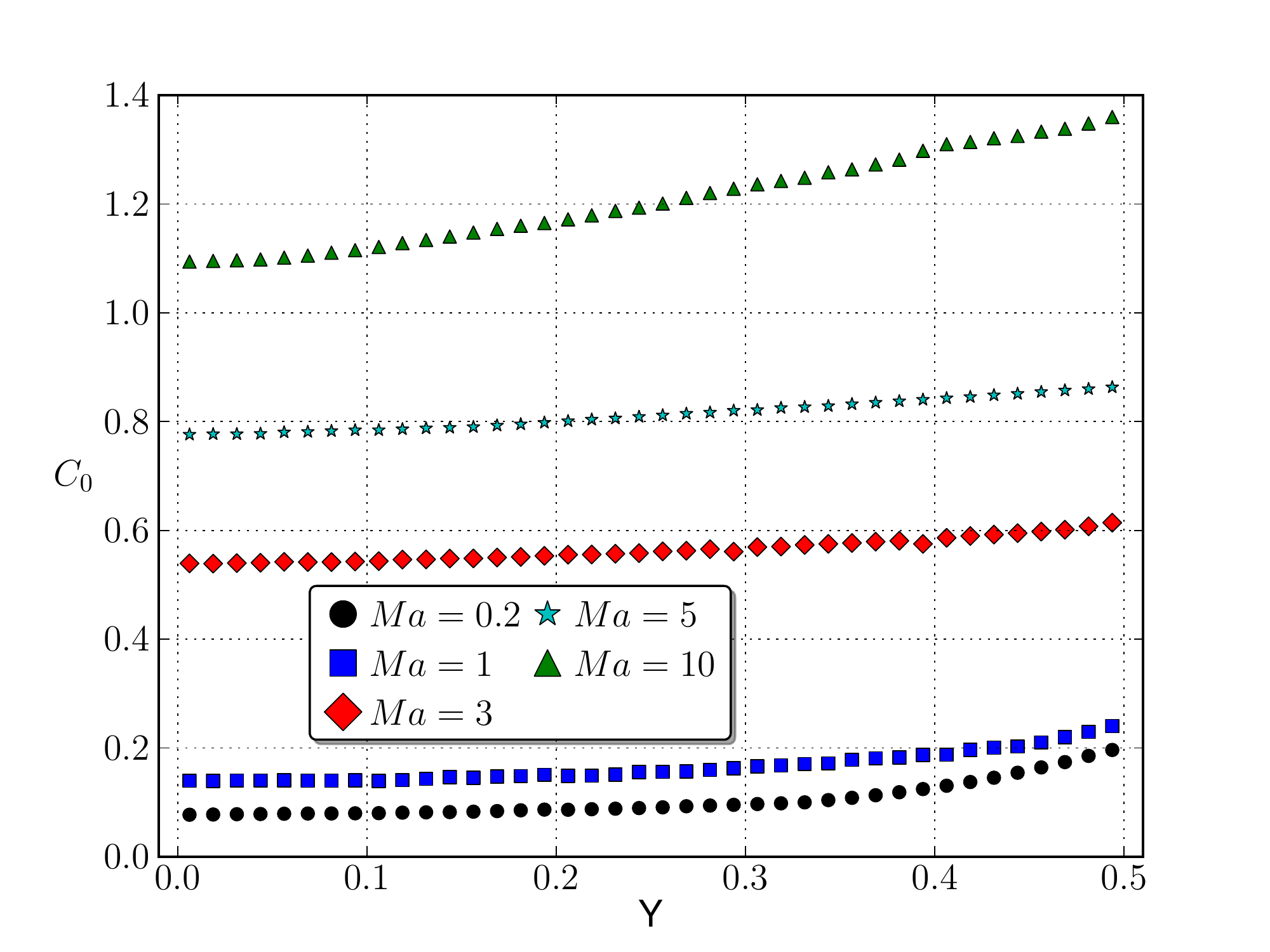}\includegraphics[width=.48
\textwidth]{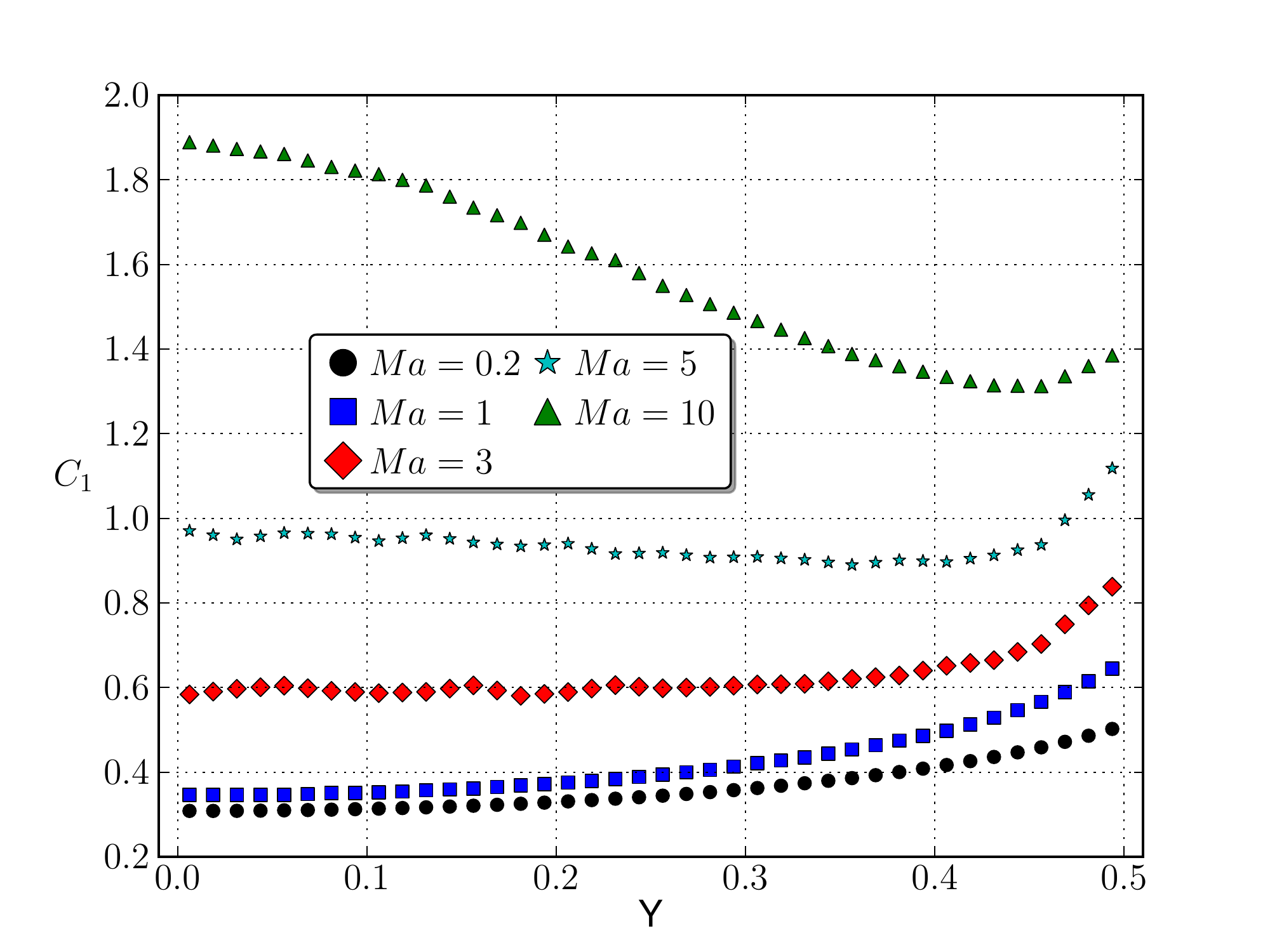}\caption{ Half-channel profiles of $C_{0}$ and $C_{1}$
at $Kn=1$, for various $Ma$.The top ($Y=0.5$) and bottom ($Y=-0.5$)
  plates are moving with speeds $U_w$ in opposite directions. \label{fig:Proc0c1kn1}}
\end{center}
\end{figure*}

The profiles of $C_{0}$ and $C_{1}$ for various Couette flow cases are presented in Figures \ref{fig:Proc0c1kn001}-\ref{fig:Proc0c1kn1}.
It is clearly shown that the level of non-equilibrium depends on both
the Knudsen number and the Mach number ($Ma=U_{w}/\sqrt{RT_0}$). $C_{0}$, the indicator of departure from equilibrium, is small at $Kn=0.01$ especially in the bulk region ($Y$=0).
However, when the Mach number increases, $C_{0}$ becomes larger.
At a higher Mach number ($Ma=5$), the values of
$C_{0}$ are even similar to those in a flow with a much higher Knudsen
number (see $Kn=0.5$ and $Ma=0.2$ in Figure \ref{fig:Proc0c1kn05}), showing
that the extent of departure from equilibrium can be similar for two
flows with very different Knudsen numbers. The Knudsen number alone does not determine the level of thermodynamic non-equilibrium.

\begin{figure*}
\begin{center}
\includegraphics[width=.48
\textwidth]{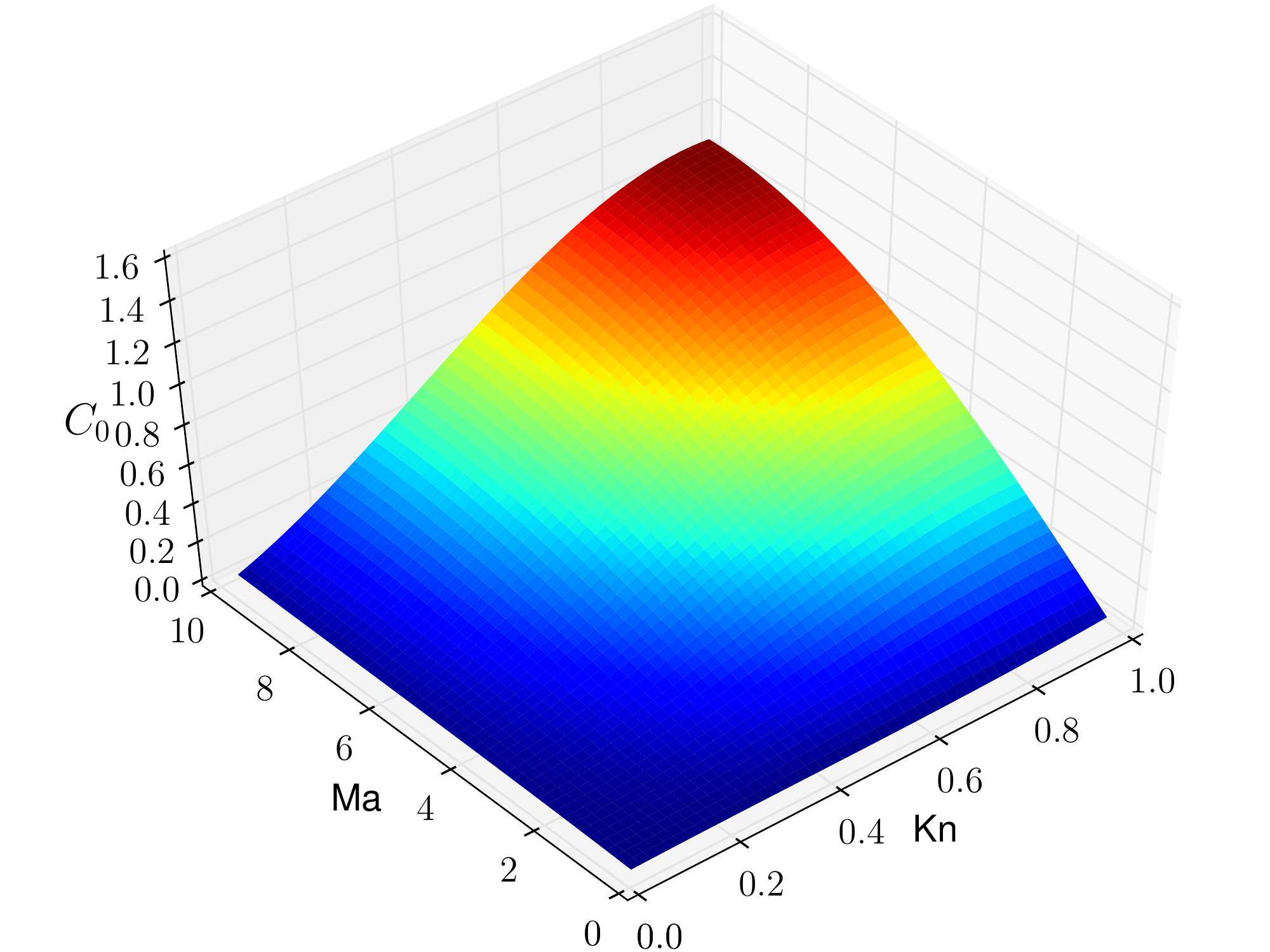}\includegraphics[width=.48
\textwidth]{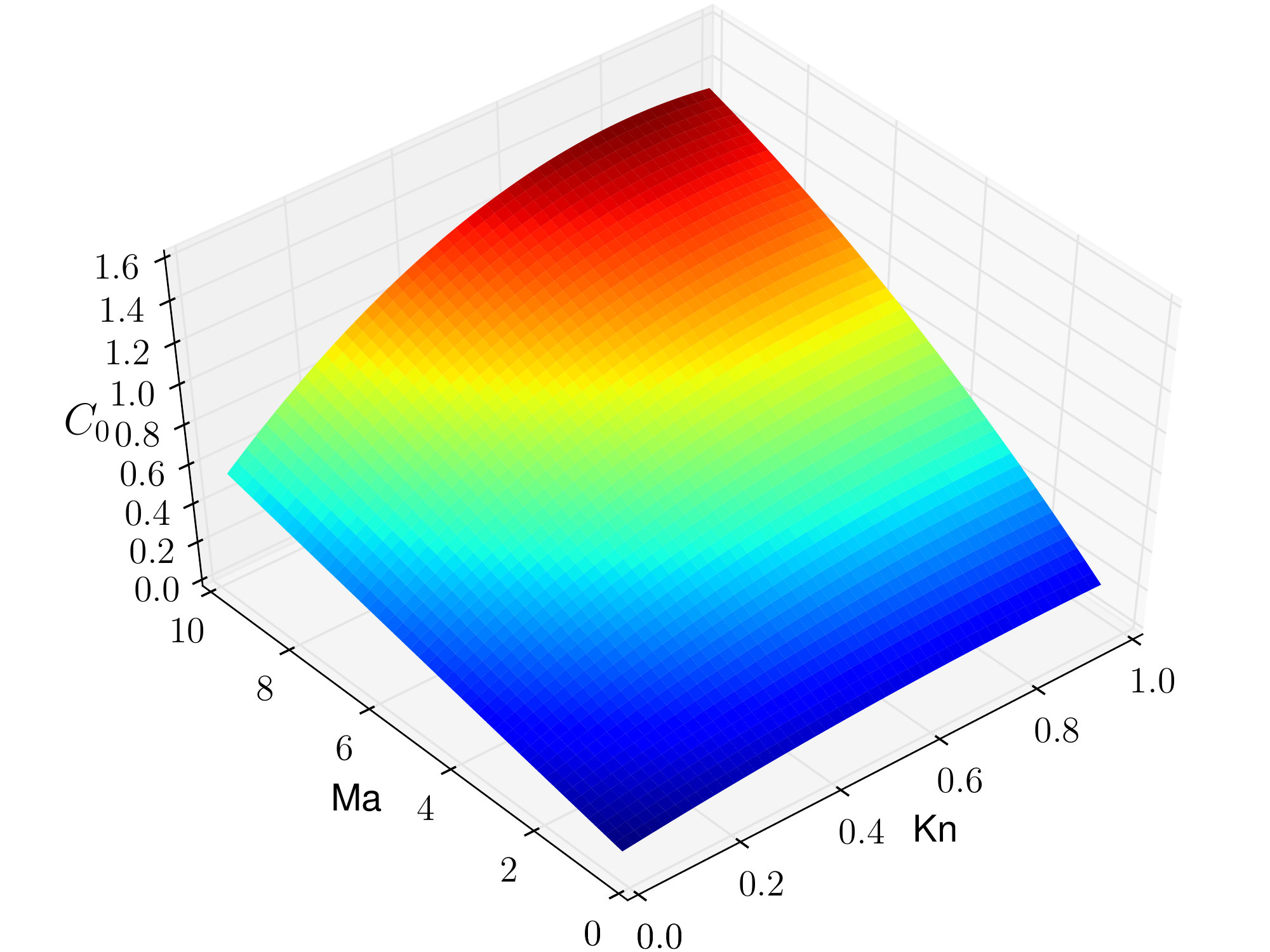}
\caption{Dependence of $C_{0}$ on $Kn$ and $Ma$ in the bulk (left)
and at the wall (right). \label{fig:c03d} }
\end{center}
\end{figure*}

\begin{figure*}
\begin{center}
\includegraphics[width=.48
\textwidth]{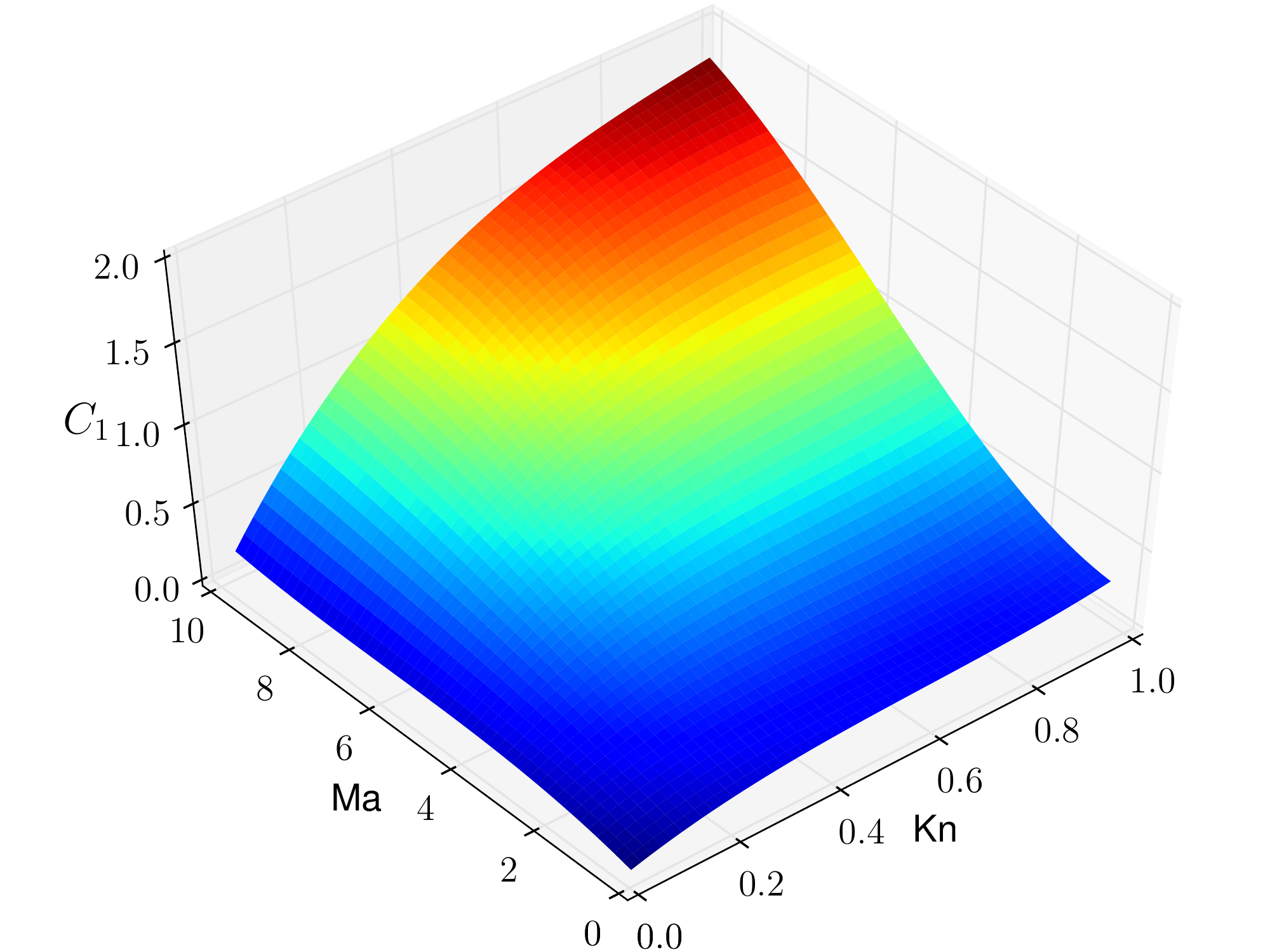}\includegraphics[width=.48
\textwidth]{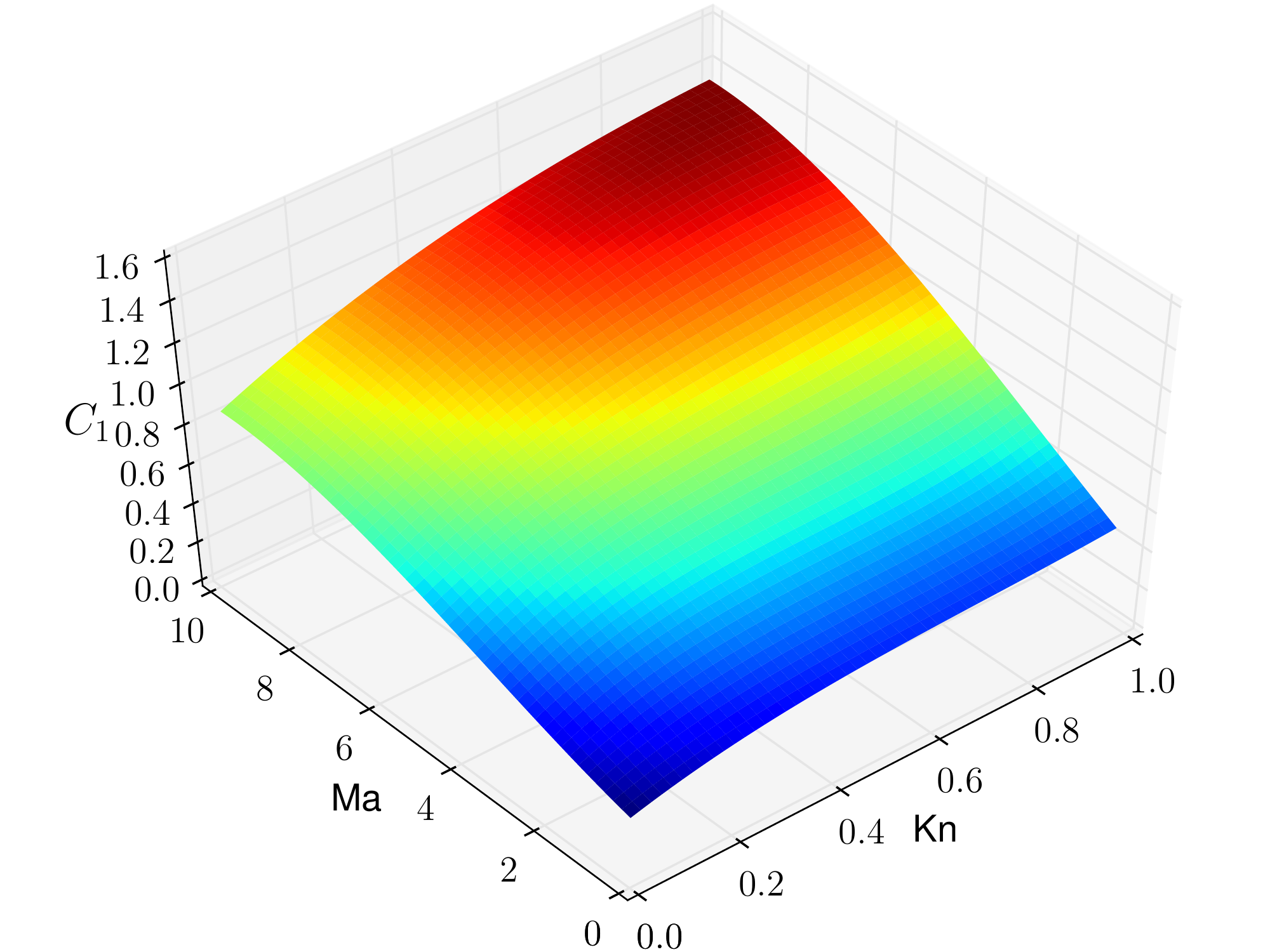}
\caption{Dependence of $C_{1}$ on $Kn$ and $Ma$ in the bulk (left)
and at the wall (right). \label{fig:c13d} }
\end{center}
\end{figure*}

For a Knudsen number of $0.01$, the Navier-Stokes equations are
usually regarded as valid. However, the value of $C_{1}$, which inversely indicates the
appropriateness of using the Navier-Stokes equations, can increase with the
Mach number, see Figure \ref{fig:Proc0c1kn001}. When the Mach number
is $1.0$ and above, a significant proportion of the non-equilibrium flow information
cannot be captured by the Navier-Stokes equations, even in this
simple flow configuration. When $Ma=0.2$, and in the bulk region ($Y=0$),
the Navier-Stokes equations hold because $C_{1}\ll1$. However, in the
wall region, the Navier-Stokes equations are less accurate. The reason
for this is the presence of Knudsen layer in the near-wall region, where the linear constitutive relation for stress that is assumed in the Navier-Stokes equations becomes inappropriate. When $Ma$ increases to $1.0$,
higher-order fluid models may be required, even in the bulk, to accurately
capture non-equilibrium information at conventional Knudsen numbers as low as 0.01. 

For flows with $Kn$ around $0.1$, it is generally assumed that the Navier-Stokes
equations will still be useful in the bulk flow region. Indeed, Figure \ref{fig:Proc0c1kn01} shows that less than $10\%$ error will be introduced to the non-equilibrium distribution function if the Navier-Stokes equations are used for a flow with $Ma=0.2$.
However, $C_{1}$ is larger than $0.1$ in the bulk region when
$Ma$ is above $0.2$, so substantial non-equilibrium flow information would
not be captured by a Navier-Stokes analysis in these cases.
Figure \ref{fig:Proc0c1kn05} also shows that non-equilibrium information
cannot be properly captured, even for Mach numbers as low as $0.2$,
when the Knudsen number becomes large (e.g. $0.5$). Figures \ref{fig:Proc0c1kn001}-\ref{fig:Proc0c1kn1} together show
that neither the Knudsen number nor the simple product of Mach and Knudsen numbers
can appropriately assess the level of thermodynamic non-equilibrium in flowfields. However, we can discover the appropriate dependencies of $C_0$ on $Kn$ and $Ma$ in certain flows.

When the Knudsen number is small, $f^{neq}\backsim f^{(1)}$
as a first order approximation,
hence,
\begin{equation}
\begin{split}
C_{0} &\sim \sqrt{\frac{\int(f^{(1)})^{2}d\bm{\xi}}{\int(f^{eq})^{2}d\bm{\xi}}}
 \\
& =\sqrt{\frac{\int(f^{eq})^{2}\left[\frac{\mu}{pRT}\frac{du<_{i}}{dx_{j>}}
\varsigma_{<i}\varsigma_{j>}+\frac{2\kappa}{5pRT}\frac{dT}{dx_{i}}\varsigma_{i}
\left(\frac{\varsigma_{j}\varsigma_{j}}{2RT}-\frac{5}{2}\right)\right]^{2}d\bm{
\xi}}{\int(f^{eq})^{2}d\bm{\xi}}}.
\end{split}
\end{equation}
For Couette flows, this can be simplified to
\begin{equation}
  C_{0}\sim 
\sqrt{\frac{\int(f^{eq})^{2}\left[\frac{\mu}{pRT}\frac{du{}_{x}}{
dy } \varsigma_{x}\varsigma_{y}+\frac{2\kappa}{5pRT}\frac{dT}{dy}\varsigma_{y}
\left(\frac{\varsigma_{x}^{2}+\varsigma_{y}^{2}+\varsigma_{z}^{2}}{2RT}-\frac{5}
{2}\right)\right]^{2}d\bm{\xi}}{\int(f^{eq})^{2}d\bm{\xi}}}.    
\label{eq:C0}
\end{equation}
In the Navier-Stokes model, the velocity gradient turns out as $2U_{w}/{L}$ over the whole flowfield. The temperature
gradient can vary with position, but is zero at the centerline
and $\thicksim 2U_{w}^{2}\mu/\kappa L$ at the wall. So
we estimate $C_{0}$ at the centerline from Eq.(\ref{eq:C0}) to be 

\begin{equation}
 C_{0}\sim\frac{U_{w}\mu}{Lp}=\frac{\mu\sqrt{RT_{0}}}{pL}Ma \backsim KnMa,
\end{equation}
and at the wall as
\begin{equation}
\begin{split}
C_{0}&\sim\sqrt{\frac{7U_{w}^{4}\mu^{2}}{10L^{2}p^{2}RT}+\frac{U_{w}^{2}\mu^{2}}
{ L^{2}p^{2}}}\\ 
&\sim\sqrt{\frac{\mu^{2}RT_{0}}{p^{2}L^{2}}\frac{T_{0}}{T}Ma^{4}
+\left(\frac{\mu\sqrt{RT_{0}}}{pL}Ma\right)^{2}}\\
& \sim KnMa\sqrt{1+Ma^{2}}.
\end{split}
\end{equation}
Therefore, the level of
non-equilibrium increases with both the Knudsen and the Mach numbers.
Note that this estimation is only appropriate when the Knudsen number
is not too large. For large $Kn$, we need to analyse the relation
numerically. For $C_{1}$, it is difficult to provide a solution 
even for small Knudsen numbers, as no generally agreed Burnett-order
solution for $f^{(2)}$ is available. The numerical data shown in Figures
\ref{fig:c03d} and \ref{fig:c13d} suggest a complicated general dependency 
of $C_{0}$ and $C_{1}$ on the Knudsen and Mach numbers. 

In conclusion, we have addressed how to accurately assess the departure of flowfields from thermodynamic equilibrium, and how to identify when the Navier-Stokes model is applicable or not. Two new kinetic parameters based on the molecular velocity distribution function are proposed, with $C_{0}$ assessing how far
flowfields are away from equilibrium, while $C_{1}$ indicating the
validity of the Navier-Stokes model. Our MD numerical experiments
conducted for Couette flows confirm that the flowfield and the appropriateness
of the Navier-Stokes equations are not properly assessed by the Knudsen number alone,
or even by the combination of Knudsen and Mach numbers proposed in the
literature. The direct indicators $C_{0}$
and $C_{1}$ show a complicated dependency on both $Kn$ and $Ma$. Our proposed
parameters are not only of theoretical interest, but could be used for model switching criteria
in kinetic-continuum hybrid numerical schemes.

\end{document}